# Estimation of excess mortality in a chronic condition from current status data with disease duration: simulation study about need for long-term care

Ralph Brinks
E-Mail: ralph.brinks@uni-wh.de

Chair for Medical Biometry and Epidemiology
Witten/Herdecke University
Faculty of Health/School of Medicine
D-58448 Witten, Germany

Institute for Biometry and Epidemiology
German Diabetes Center
D-40225 Düsseldorf, Germany

## Abstract

This article describes a method to estimate the mortality rate ratio $R$ from current status data with duration in a chronic condition in case the general mortality of the overall population is known. Apart from the general mortality, the method requires four pieces of information from the study participants: age and time at the survey/interview, whether the chronic condition is present (current status) and if so, for how long the condition is present (duration). The method uses a differential equation that relates prevalence, incidence and mortality to estimate $R$ of the people with the chronic condition compared to those without the condition.

To demonstrate feasibility, a simulation based on the illness-death model (multi-state model) with transition rates motivated from long-term care is run. It is found that the method requires a large number of study participants (100000 or more) to estimate $R$ with a reasonably low relative error. Despite the large sample size, the method can be useful in settings when cross-sectional information are easily available, e.g., in claims data, and national age-specific general mortality rates are accessible from vital statistics.

## Introduction

In a review paper from 1991, Keiding has raised the question of inference from current age and current duration in a cross-sectional sample [Kei91]. Keiding was referring to the illness-death model for irreversible diseases shown in Figure 1, where the population under consideration is divided into the states *Healthy, Diseased* and *Dead*. The transition rates (or, synonymously, intensities) between the states are denoted by $\lambda$, $\mu_0$, and $\mu_1$.

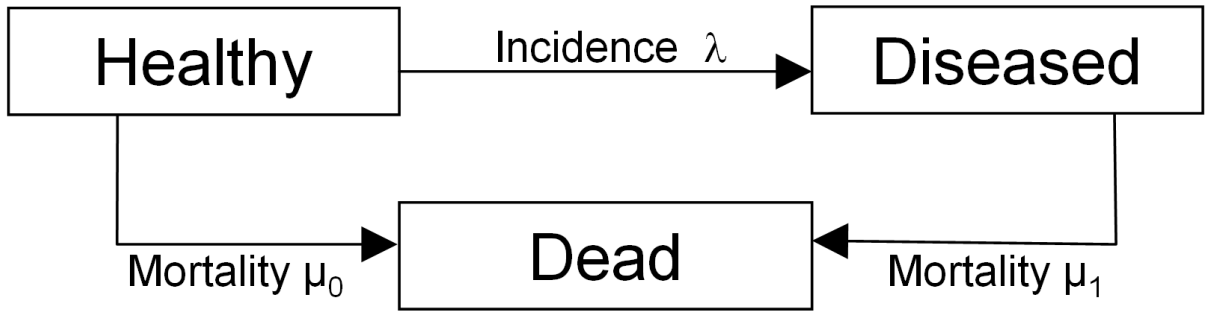

**Figure 1: Illness-death model for a chronic condition ('Diseased') and associated transition rates: incidence $\lambda$, mortality without ($\mu_0$) and with the disease ($\mu_1$).**

The three rates $\lambda$, $\mu_0$, and $\mu_1$ may depend on calendar time $t$ and on age $a$. In epidemiological contexts, calendar time is frequently called period. By current age and current duration of an individual $i$, it is meant the current age $a_i$ at survey time $t_i$ for healthy as well as diseased individuals, disease status $\delta_i = 1$ if and only if subject $i$ is diseased at age $a_i$, and $D_i$ the duration since the onset of the disease. In case subject $i$ is not diseased, $D_i$ is set to be 0. For each individual $i$ surveyed at time $t_i$, we have the information represented by the quadruple

$$(t_i, a_i, \delta_i, D_i), \qquad i = 1, ..., n. \qquad (1)$$

Apart from the quadruples (1), in this paper we assume that the general mortality $\mu$ in the population is known. Frequently, these are recorded at the national statistical offices. In this paper, it is shown that these information can be used to estimate the mortality rate ratio (MRR) $R = \mu_1/\mu_0$.

## Illness-death model and partial differential equation

Recently, it has been shown that the age-specific prevalence $\pi$ of a chronic condition at some time $t$ is related to the rates $\lambda$, $\mu_0$, and $\mu_1$ via a partial differential equation (PDE) [Bri14]. We consider the case without migration, where the PDE from [Bri14] reads as

$$(\partial\pi/\partial t + \partial\pi/\partial a) = (1 - \pi) \times [\lambda - \pi (\mu_1 - \mu_0)]. \qquad (2)$$

Using the relations $R = \mu_1/\mu_0$ and $\mu = \pi \mu_1 + (1 - \pi) \mu_0 = \mu_0 [\pi R + (1 - \pi)]$ with $\mu$ being the general mortality in the population (sometimes known as the overall mortality), Equation (2) becomes

$$(\partial\pi/\partial t + \partial\pi/\partial a) = (1 - \pi) \times [\lambda - \mu \pi (R - 1)/\{1 + \pi (R - 1)\}]. \qquad (3)$$

Equation (2) is the basis for an estimation strategy for the quantity *R* from the quadruples (1) given that the general mortality μ is known. If π, λ and μ are known, Equation (3) can be solved for the MRR *R*:

$$R = 1 + \{\lambda (1 - \pi) - \partial \pi\} / \{\pi (1 - \pi)(\mu - \lambda) + \pi \, \partial \pi\}, \qquad (4)$$

where, for brevity, we set $\partial \pi := \partial \pi / \partial t + \partial \pi / \partial a$.

The advantage of Equation (4) lies in the fact that incidence and prevalence can directly be estimated from the quadruples (1), which can be obtained from (repeated) cross-sections, without the need to follow-up study participants. Following-up study participants is time-consuming and usually requires administrative efforts. The general mortality μ in Equation (4) can often be obtained from official vital statistics. Many countries provide vital statistics of the overall population in regions by their national statistical offices. In case a region or country does not offer these information, scientific databases like the Human Mortality Database (mortality.org) or mortality databases from global organizations like the United Nations' Population Division may help.

## Estimation of the mortality rate ratio

In the previous section, we have seen how Equation (4) leads us to a way of estimating the MRR $R = \mu_1/\mu_0$ from the quadruples in (1) if the general mortality μ is given. This gives rise to

Algorithm 1:
(i)   Estimate the incidence rate λ(*t*, *a*) from the quadruples (1).
(ii)  Estimate the prevalence π(*t*, *a*) from the quadruples (1) and estimate ∂π(*t*, *a*).
(iii) Plug-in the estimates from steps (i) and (ii) into Equation (4) to estimate the MRR *R*.

The three steps (i) to (iii) of Algorithm 1 need some clarification. Estimation of the incidence rate λ as a function of two time scales, calendar time *t* and age *a*, is frequently done by a Poisson regression after a Lexis expansion. Details are given in the textbooks [Cla93, Car21] or in the article [Ber94]. The latter reference has a more Bayesian focus. Estimation of the prevalence π at (time, age) = (*t*, *a*) can be accomplished by estimating the number of people with disease at (*t*, *a*) divided by the number of people alive at (*t*, *a*). Usually, this is done similar to a Lexis expansion by counting people with disease and people alive in (*t*, *a*)-bands with a specified width *w*, say *w* = 1 year.

## Simulation: need of long-term care

In this example, we estimate the MRR in a simulated elderly population using the algorithm in the previous section. For this, we first simulate a population ℘ of 500000 subjects moving through the illness-death model in Figure 2. The subjects in the population ℘ are followed from birth to death. Hence, there is no censoring. The age-specific incidence and mortality rates are motivated by the need for long-term care in Germany. We consider the *Diseased* state in Figure 1 as 'Need for long-term care'. This state is considered to be irreversible: Once someone has entered this state, the person cannot go back to the *Healthy* state. The corresponding transition rates in the illness-death model are assumed to be exponential in terms of age *a* (see Table 1) and are shown in logarithmic plots in Figure 2.

| Rate | Functional form | Coefficients | |
|---|---|---|---|
| Incidence | $\lambda(a) = \exp(\beta_0 + \beta_1 \times a)$ | $\beta_0 = -9.5$ | $\beta_1 = 0.085$ |
| Mortality without disease | $\mu_0(a) = \exp(\alpha_{00} + \alpha_{01} \times a)$ | $\alpha_{00} = -11$ | $\alpha_{01} = 0.11$ |
| Mortality with disease | $\mu_1(a) = \exp(\alpha_{10} + \alpha_{11} \times a)$ | $\alpha_{10} = -9.5$ | $\alpha_{11} = 0.095$ |

**Table 1: Input values for the rates in the simulation.**

The left panel of Figure 2 shows the age-specific incidence rate $\lambda$, the right panel shows the mortality rates $\mu_0$ (solid line) and $\mu_1$ (dashed line) without and with need for long-term care, respectively.

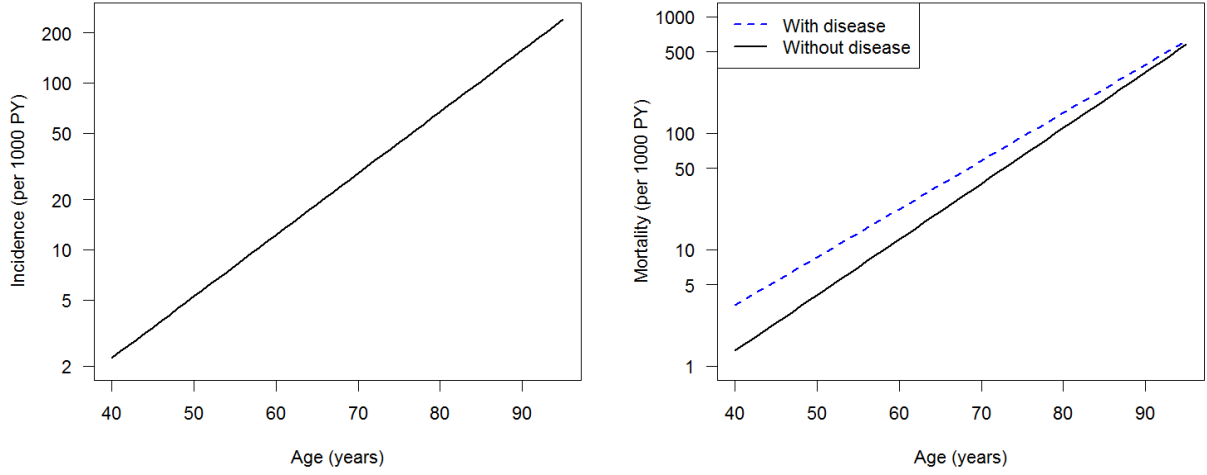

**Figure 2: Age-specific incidence $\lambda$ (left) and mortality rates $\mu_0$, $\mu_1$ (right).**

The simulation of the people in the population $\wp$ moving through the illness-death model is done by the algorithm presented in [Bri14a]. In short: for each subject in $\wp$, the ages at which transitions in the illness-death model take place are randomly drawn based on the rates shown in Figure 2 by a Monte Carlo simulation. For each simulated subject, two situations may occur: If a subject is assigned to the *Diseased* state according to rate $\lambda$, a second event (death after being *Diseased*) with associated rate $\mu_1$ is simulated. Alternatively, subjects may die with rate $\mu_0$ without being in the *Diseased* state.

After each subject's path in the illness-death model is simulated, we mimic a survey to obtain the quadruples as in Eq. (1). For the survey, we first randomly draw $n$ survey times $t_i$, $i = 1, ..., n$, at which subjects are surveyed. Then, second, for each of the survey times $t_i$, we randomly draw a study participant from all those subjects of $\wp$ who are still alive at $t_i$. For each of the selected subjects, we calculate the values $a_i$, $\delta_i$, and $D_i$ yielding the required collection of quadruples. Based on the collected quadruples, Algorithm 1 is applied and the mortality rate ratio is estimated. To assess the effect of the sample size $n$, i.e., the number of collected quadruples, we use sample sizes $n = 5000, 10000, 20000, 50000, 100000, 200000$.

The workflow is summarized in Figure 3.

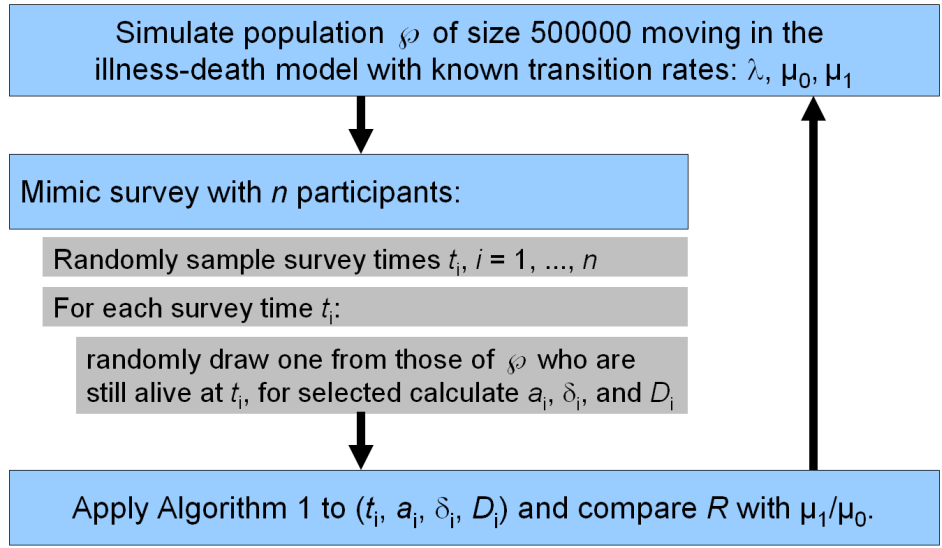

Figure 3: Workflow of the simulation. The simulation

To check the results of the simulation, we compare the empirical age-specific prevalence in the population $\wp$ with the theoretical prevalence from solving the PDE (2). For solving Eq. (2), we note that by setting $y = \pi/(1 - \pi)$, Eq. (2) becomes the linear PDC $\partial y = \lambda - y\,(\mu_1 - \mu_0 - \lambda)$, which can be solved analytically for given rates $\mu_1$, $\mu_0$, and $\lambda$ [Bru99]. Back-transformation of $y$ yield the prevalence $\pi = y / (1 + y)$. The result of the comparison between the analytically calculated prevalence and the empirical prevalence (estimated with $w = 1$ year) in the population $\wp$ is shown in Figure 4.

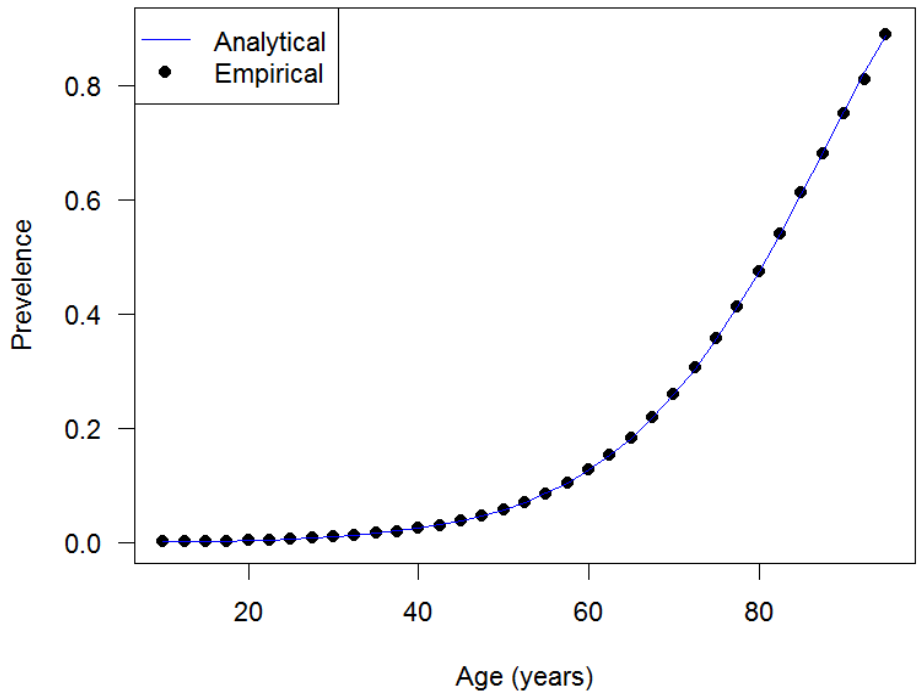

Figure 4: Analytically calculated age-specific prevalence (blue line) and the empirical prevalence in the population $\wp$ (black dots).

Given the quadruples $(t_i, a_i, \delta_i, D_i)$, $i = 1, ..., n$, we apply a Lexis expansion followed by a Poisson regression to estimate $\lambda(a)$ with age $a$ as independent variable. Table 2 shows the results of the Poisson regression depending on the number of people $n$ in the simulated survey. By increasing sample size *n*, we see how the standard errors of both, intercept $\beta_0$ and coefficient $\beta_1$, decreases.

|  | Intercept $\beta_0$ | | Coefficient $\beta_1$ | |
|---|---|---|---|---|
| Sample size *n* | Estimate | Standard error | Estimate | Standard error |
| 5000 | -9.36188 | 0.12963 | 0.08219 | 0.00197 |
| 10000 | -9.42951 | 0.09299 | 0.08331 | 0.00141 |
| 20000 | -9.54052 | 0.06721 | 0.08483 | 0.00101 |
| 50000 | -9.53298 | 0.04242 | 0.08472 | 0.00064 |
| 100000 | -9.51751 | 0.02988 | 0.08454 | 0.00045 |
| 200000 | -9.49430 | 0.02096 | 0.08429 | 0.00032 |

**Table 2: Estimated coefficients from the Poisson regression.**

Estimating the age-specific prevalence $\pi$ for the different sample sizes *n* is accomplished by dividing the number of people with disease by the number people alive in age-bands with *w* = 1 year. Then, a smoothing spline of degree 3 is fit through the estimated data points. Figure 5 shows the estimated data points (filled black circles) and the modeled equivalents (open circles). For comparison, the true (i.e., analytically calculated) age-specific prevalence is shown as solid blue line in Figure 5. The fit by the smoothing spline is done to have a robust estimate for the derivative $\partial \pi$ available, which is needed in applying Eq. (4).

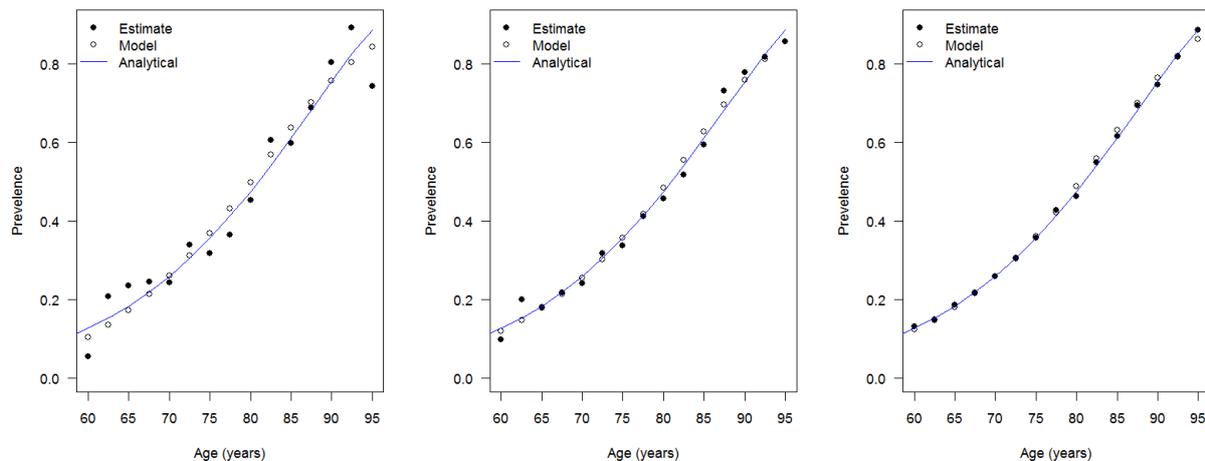

**Figure 5: Estimated (filled black circles), modeled (open circles) and true prevalence (blue line) for *n* = 5000 (left panel), *n* = 20000 (middle panel), and *n* = 200000 (right panel).**

With the estimated incidence $\lambda$ and prevalence $\pi$, we can use Eq. (4) to estimate the mortality rate ratio *R*. In order to get an impression, which estimation steps (estimation of incidence or prevalence), are crucial, we first assume that the prevalence $\pi$ is known and $\lambda$ is estimated, second, we assume that $\lambda$ is known and $\pi$ is estimated. Finally, $\pi$ and $\lambda$ are estimated (which reflects epidemiological practice). By known values, the values used as input for the simulation ($\lambda$ as in Figure 2) or analytically calculated ($\pi$ as shown in solid blue line in Figure 4) are meant. The numerical results are shown in Table 3. In case of *n* = 5000 and *n* = 200000 the results are also graphically displayed in Figure 6.

| Estimated | Age | True R | Estimated R with sample size n = | | | | | |
|---|---|---|---|---|---|---|---|---|
| | | | 5000 | 10000 | 20000 | 50000 | 100000 | 200000 |
| λ | 65 | **1.690** | 1.464 | 1.489 | 1.430 | 1.432 | 1.448 | 1.484 |
| | 70 | **1.568** | 1.381 | 1.413 | 1.399 | 1.399 | 1.406 | 1.425 |
| | 75 | **1.455** | 1.306 | 1.337 | 1.343 | 1.342 | 1.345 | 1.355 |
| | 80 | **1.350** | 1.233 | 1.260 | 1.274 | 1.272 | 1.273 | 1.278 |
| | 85 | **1.252** | 1.160 | 1.184 | 1.200 | 1.198 | 1.198 | 1.200 |
| | 90 | **1.162** | 1.090 | 1.109 | 1.125 | 1.123 | 1.123 | 1.123 |
| | 95 | **1.078** | 1.020 | 1.036 | 1.051 | 1.050 | 1.049 | 1.049 |
| π, ∂π | 65 | **1.690** | 0.896 | 1.586 | 1.658 | 1.476 | 1.712 | 1.656 |
| | 70 | **1.568** | 1.221 | 1.412 | 1.525 | 1.342 | 1.460 | 1.469 |
| | 75 | **1.455** | 1.253 | 1.268 | 1.356 | 1.273 | 1.300 | 1.319 |
| | 80 | **1.350** | 1.204 | 1.203 | 1.242 | 1.225 | 1.219 | 1.232 |
| | 85 | **1.252** | 1.238 | 1.212 | 1.213 | 1.216 | 1.205 | 1.208 |
| | 90 | **1.162** | 1.317 | 1.275 | 1.258 | 1.254 | 1.240 | 1.236 |
| | 95 | **1.078** | 1.389 | 1.334 | 1.318 | 1.312 | 1.298 | 1.292 |
| λ, π, ∂π | 65 | **1.690** | 0.713 | 1.383 | 1.393 | 1.213 | 1.460 | 1.449 |
| | 70 | **1.568** | 1.062 | 1.266 | 1.357 | 1.184 | 1.303 | 1.332 |
| | 75 | **1.455** | 1.125 | 1.164 | 1.250 | 1.172 | 1.200 | 1.228 |
| | 80 | **1.350** | 1.104 | 1.126 | 1.173 | 1.157 | 1.152 | 1.169 |
| | 85 | **1.252** | 1.151 | 1.148 | 1.163 | 1.165 | 1.154 | 1.160 |
| | 90 | **1.162** | 1.229 | 1.214 | 1.216 | 1.211 | 1.197 | 1.194 |
| | 95 | **1.078** | 1.296 | 1.272 | 1.278 | 1.271 | 1.257 | 1.251 |

**Table 3: Estimated and true age-specific mortality rate ratios R for different ages and sample sizes n.**

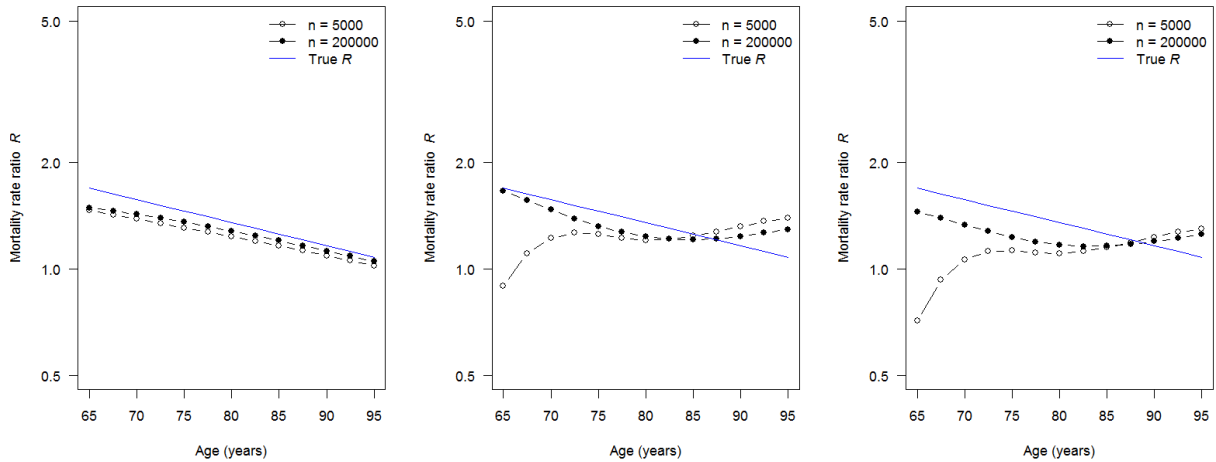

**Figure 6: Estimated age-specific mortality rate ratio R for sample sizes n = 5000 (open circles) and n = 200000 (filled circles). The panels shows the situations where λ (left panel), π (mid panel), and λ, π (right panel) are estimated from the data. For comparison, the R values used as input for the simulation ('True R') is shown as blue solid line.**

From Table 3 and Figure 6 we see that the estimated prevalence π mostly imposes the biggest error to the estimate of R, irrespective of the sample size n. As expected, with increasing sample size n the estimated mortality rate ratios R approach the R values used as input for the simulation (true R).

# Discussion

In this work, we have described a method to estimate the mortality rate ratio $R$ in a chronic condition from four pieces of information of study participants: age and time at the interview, whether the chronic condition is present and if so, for how long. Given the mortality in the overall population, these four pieces of information allow us to estimate the age-specific prevalence and incidence of the chronic condition. This possible by using a differential equation that relates prevalence, incidence and mortality to estimate $R$ of the people with the chronic condition compared to those without the condition. A prerequisite of the method is that the general mortality rate in the overall population is know. For many countries, the general mortality is well-known from the national offices for statistics like, e.g., the US Census Bureau or the UK Office for National Statistics. Given the general mortality, the new method does not require follow-up of study participants but only repeated cross-sectional information. This can be seen as an advantage, because following study participants up is costly and time consuming. Another advantage lies in the fact that it is not necessary to rely on any information about cause of death or if the chronic condition is present at the time of death.

As demonstrated above, the method requires a large number of study participants, In the given example, 100000 or more study participants are necessary to estimate $R$ with a relative error consistently below 20%. That a large numbers of study participants are necessary to have reasonably accurate estimates for the prevalence and incidence has also been observed in a slightly different setting [Bri21].

The question arises if there are situations where the method is useful. The example about people with need for long-term care has been chosen on purpose. In Germany, for instance, people in need for long-term care are mandatorily registered at a speical insurance, which allows to collect the necessary cross-sectional information about presence of need for long-term care and duration. Moreover, the German Federal Statistical Office provides the national age-specific general mortality rate. Complemented with the age distribution of Germany, these information would allow to estimate the age-specific mortality rate ratio $R$ of people with need for long-term care. This is considered useful, because for Germany, so far, there are no estimates about mortality of people with need of long-term need compared to those without this need.

# Additional information

## *Availability of source code*

The source code for running the simulation with the statistical software R (The R Foundation for Statistical Computing) is available in the public repository *Zenodo* under digital object identifier (DOI) 10.5281/zenodo.7227909.

## *Funding statement*

The author did not receive any funding for any aspect of this work.

## *Competing interests*

The author declares that no competing interests exist with any aspect of this work.

# References

[Ber94] Berzuini C, Clayton D. Bayesian analysis of survival on multiple time scales. Statistics in Medicine, 13(8): 823-838, 1994

[Bri14] Brinks R, Landwehr S: Age-and time-dependent model of the prevalence of non-communicable diseases and application to dementia in Germany. Theoretical Population Biology, 92:62-68, 2014

[Bri14a] Brinks R, Landwehr S, Fischer-Betz R, Schneider M, Giani G. Lexis Diagram and Illness-Death Model: Simulating Populations in Chronic Disease Epidemiology. PLOS ONE 9(9): e106043, 2014

[Bri21] Brinks R, Tönnies, Hoyer A. Impact of diagnostic accuracy on the estimation of excess mortality from incidence and prevalence: simulation study and application to diabetes in German men. F1000Research 10:49, 2021

[Bru99] Brunet RC, Struchiner CJ: A non-parametric method for the reconstruction of age- and time-dependent incidence from the prevalence data of irreversible diseases with differential mortality. Theoretical Population Biology 56(1): 76-90, 1999.

[Car21] Carstensen B, Epidemiology with R, Oxford University Press, 2021

[Kei91] Keiding N: Age-specific incidence and prevalence: a statistical perspective. J Royal Stat Soc A: 154 (3), 371-412, 1991